# Metastatic melanoma-A review of current and future perspective

Qurat-ul-Ain,[1, 2*] and Muhammad Ismail[3]

[1]Dr. Panjwani Center for Molecular Medicine and Drug Research, International Center for Chemical and Biological Sciences, University of Karachi, Karachi-75270, Pakistan
[2]H. E. J. Research Institute of Chemistry, International Center for Chemical and Biological Sciences, University of Karachi, Karachi-75270, Pakistan
Helmholtz Zentrum München, Institute of Metabolism and Cell Death, Neuherberg, Germany

## ABSTRACT

Metastatic Melanoma, the fifth most common cancer in the western countries and the most common malignancy diagnosed in United States present itself as the most lethal treatment resistant cancer worldwide. In addition to the reactive oxygen species (ROS), mutations in the genes encoding receptors and non-receptor tyrosine/serene/threonine protein kinases are known to be involved in its etiology. Kinases are molecular players of cell survival, growth, and proliferation and migration that mediate their effects *via* various signal transduction pathways. A number of such molecular players have been previously found to be mutated and hyper phosphorylated in melanoma. Although, several systemic therapies including cytotoxic chemotherapy, targeted drugs, hormonal therapy, radiation therapy, bio-chemotherapy, and therapies that inhibit negative regulation of immune system have been approved from U. S. Food and Drug Administration (FDA) for metastatic melanoma treatment. However, no systemic therapy has meaningfully changed its survival end points so far and surgery still presents primary treatment option for advanced and metastatic melanomaa due to its highly resistant nature towards systemic drugs, high rate of severe, life-threatening, or fatal side effects, and un satisfactory overall response rate. Therefore, there is still a need to develop therapies that target the unique molecular profile of melanoma tumors.

Contents

*Corresponding Author:Quratulain@iccs.edu,quratulain393yahoo.com, Tel.: 03363668337

**Basic Notions of Melanocytes and Melanoma Biology**

Melanoma originates from melanocytes. It is one of the most dangerous, drug-resistant, and highly metastatic skin cancers (Bandarchi, Jabbari, Vedadi, and Navab, 2013). The melanocytes transform into melanoma cells under the influence of various chemical, environmental, and genetic factors. ROS act on melanocytes as they are released from their primary stem cell pool, and also act during their proliferation (Bisevac *et al*., 2018; Dayem *et al*., 2018). Genome-wide sequencing approaches have revealed that melanoma includes a complex of thousands of "passenger" and "driver" mutations comprising changes in DNA methylation, deletions, translocations, and amplification (Bandarchi, Navab, Seth, and Rasty, 2010). Approximately 50% of the transformed cells contain mutation in the BRAF gene that encodes a threonine/ serine kinase, which is a component of the MAP kinase pathway that controls cell growth (Ho, Jonsson, and Tsao, 2017). BRAF mutation results in hyperphosphorylation of MAP kinases activity, subsequently leading to uncontrolled growth of melanoma cells (Pons, and Quintanilla, 2006). However, together with BRAF mutation, melanocytes also undergo mutations in the genes of two important cell cycle regulatory proteins, p16$^{INK4A}$ and p14$^{ARF}$, to acquire full cancerous phenotype. Both of these proteins, encoded by CDKN2A gene, are found to be deleted in almost 16–41% of sporadic melanomas, constituting 2$^{nd}$ highest penetrance genetic alteration. Hence, deletion, promoter silencing or mutation in CDKN2A gene inactivates its function and leads to unregulated cell growth. The CDK4 gene mutation is the third most important high penetrance mutation. Proteins encoded with CDKN2A and CDK4 regulate cell cycle. Therefore, mutation in these genes disturbs the G1/S-phase checkpoint with a subsequent dysregulated cell growth (Sanches, Almeida, and Freitas, 2018). All these genetic and environmental changes are considered important for the transformation of melanocytes from healthy phenotype to tumor phenotype (Satyamoorthy and Herlyn, 2002).

**Melanoma, Survival, and Proliferation**

In the later stage of melanoma progression, the genetically and environmentally modified genes that encode kinases become part of a pathway that controls melanoma survival and proliferation. Indeed, in most melanomas, the CDK4 pathway is dysregulated either due to the loss of p16INK4A gene, or as a consequence of ERK hyerphosphorylation that initiates the RAS–RAF–MEK–ERK pathway, and stimulates cell cycle progression. The latter is an important contributor to melanoma evolution (Xu and McArthur, 2016). However, identification of ideal pathways to be used as therapeutical targets for maximal clinical benefit in melanoma patients remains a challenge.

**The Kinases as Molecular Players of Melanoma Cell Survival, Proliferation, and Metastasis**

The enzymes kinases responsible for phosphorylation reactions are either transmembrane receptors or cytosolic proteins. There are 518 putative kinases in the human genome, out of which 90 are "tyrosine kinases" (PTKs) and 428 are "serine/threonine kinases" (PSKs) (Angus, Zawistowski, and Johnson, 2018). In signal transduction pathways, kinases catalyze phosphorylation reactions in a well-orchestrated manner. The deregulation of this cascade leads to changes in survival, proliferation, and migration of normal body cells, and results in various ailments, including many cancer types (Nishi, Shaytan, and Panchenko, 2014).

Phosphorylation, is the most abundant post-translational modification in eukaryotic proteins, it is the process of addition of phosphate ($PO_4^{3-}$) group to a protein, with consequent conformational changes in its structure. The hydrophobic group of an amino acid residue, when esterified with phosphate ($PO_4^{3-}$) can turn a hydrophobic part of a protein into a polar one. The consequent changes of protein dynamics *via* distant allosteric interactions within the protein cause proteins to become activated or deactivated (Kemp, 2018). Although lipid, sugars, and other biological macromolecules have a tendency to give up or take-up ($PO_4^{3-}$) groups, only proteins are biological macromolecules that are abundantly phosphorylated as between one third to two thirds of the proteome is found to be temporarily phosphorylated in eukaryotes (Sacco, Perfetto, and Cesareni, 2017).

Tyrosine, serine, threonine, and histidine residues of proteins are subjected to phosphorylation in eukaryotes whereas there are some other amino acid residues that become phosphorylated in prokaryotes (Grangeasse, Stülke, and Mijakovic, 2015; Krebs and Beavo, 1979; Cain, Solis, and Cordwell, 2014). Serine is the most commonly phosphorylated residue, followed by threonine. Tyrosine phosphorylation is rare in most eukaryotes. However, phosphorylation of tyrosine residues is the key step in many protein phosphorylation signaling pathways (e.g., in tyrosine kinase-linked receptors) (Pearson and Kemp, 1991). Phosphorylation and dephosphorylation are critical for regulating the functions of many cellular processes. Although phosphorylation regulates many biochemical and pathophysiological processes, it is also involved in the thermodynamics of energy-requiring reactions, protein-protein interactions *via* recognition domains, and enzyme inhibition, and also plays a role in protein degradation (Bononi *et al*., 2011; Cheng, Qi, Paudel, and Zhu, 2011). All these events occur by switching on and off various signal transduction pathways of normal cell metabolism, transcription, cell cycle progression, cell survival, rearrangement of cytoskeletan, cell migration, apoptosis, and cell proliferation (Day, Sosale, and Lazzara, 2016; Nishi, Shaytan, and Panchenko, 2014).

**Tyrosine Specific Protein Kinases**

These enzymes are either present in the cell as cytosolic enzymes or on plasma membranes as receptors to phosphorylate only, the

## Receptors Tyrosine Kinase
### Platelet Derived Growth Factor Receptor (PDGFR)

PDGF receptors are cell surface kinase receptors that are activated with the binding of their cognate ligand, the platelet derived growth factor (PDGF). Activated PDGFR dimerizes, and activates various intracellular cytosolic kinases by autophosphorylation. subsequently, signal transduction pathways are activated, for instance, PI3K pathway, MAP kinase pathway mTOR, and SFKS pathway (Fig. 1.1). All these pathways ultimately regulate cell cycle, cell growth, survival, cell prolifearion and cell migration by effecting gene expression. PDGFRs have been identified as two types that bind with different PDGF; PDGF-BB, and PDGF-AB (Kumar *et al.*, 2017).

Tyrosine 751 phosphorylation site of PDGF-β that resides in the cytosol, autophosphorylates when the receptor binds with its cognate ligand. This tyrosine 751 residue binds with PI3K with its regulatory subunit p85. Upon tyrosine 751 phosphorylation, PI3K becomes activated, and converts PIP2 into PIP3, enhancing the PI3-AKT pro survival pathway for a normal cell growth (Szöőr *et al.*, 2016).

### Epidermal Growth Factor Receptor (EGFR)

EGF-receptor is a transmembrane protein. Epidermal growth factor (EGF), and transforming growth factor α (TGFα) are specific ligands of (EGFR) that bind with EGFR and activate it. Upon ligand binding, EGFR forms an active homodimer from inactive monomeric form (Fig. 1.1) (Wee and Wang, 2017).

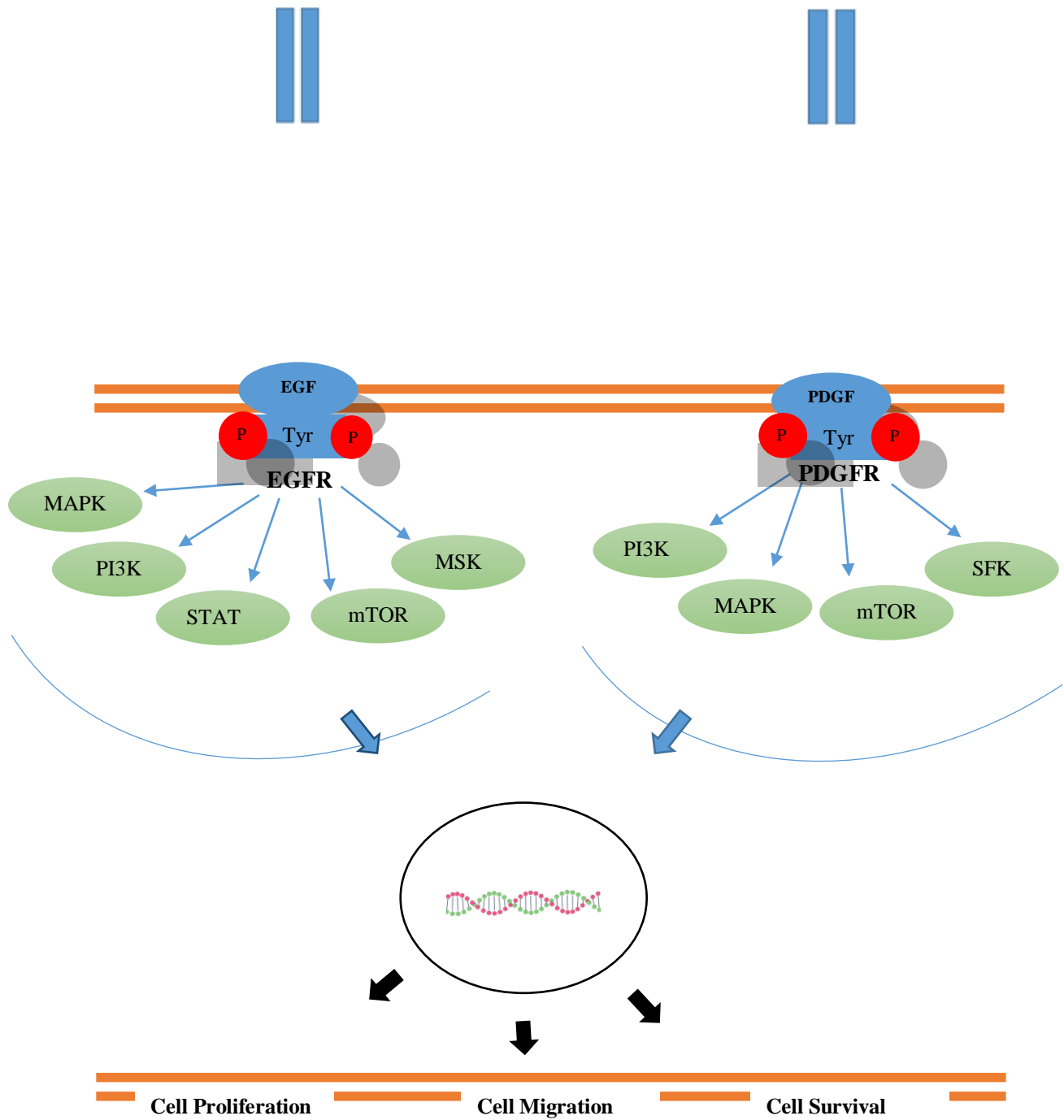

Dimerization stimulates intrinsic intracellular protein-tyrosine kinase activity with subsequent auto-phosphorylation of several tyrosine (Y) residues (Addison, 2014). When tyrosine residues, including Y992, Y1045, Y1068, Y1086, Y1148, and Y1173 of the C-terminal EGFR auto-phosphorylate, they serve as docking sites for other downstream proteins of several signal transduction cascades, principally the MAPK, STAT, PI3K, and MSK mTOR pathways, leading to DNA synthesis, cell migration, adhesion, and proliferation (Wee and Wang, 2017).

**Non-receptors Tyrosine Kinase**
*The Src Family of Protein Tyrosine Kinases (SFKs)*

These are membrane-associated non receptor tyrosine kinases with proto-oncogen and signal transduction regulatory activity. This family of kinases comprises of thirteen members, including: Yes, Fyn, Fgr, c-Src, Yrk, Frk, Lyn, Blk, Hck, FAK, and Lck, all of them regulate signal transduction pathways through various cell surface receptors: adhesion receptors, receptor tyrosine kinases, G-protein coupled receptors, and cytokine receptors in the context of multiple cellular environments (Espada and Martin-Perez, 2017; Vlaeminck-Guillem, Gillet, and Rimokh, 2014). Most studies are based on the PDGFR and EGFR tyrosine kinases receptors (Lewis-Tuffin *et al.*, 2015). When SFKs are stimulated by these receptors, they phosphorylate specific tyrosine residues in other proteins involved in cell survival, cell growth, invasion, migration, and metastatic signal transduction pathways. Therefore, these kinases act as important intermediaries of signal transduction pathways of both cancerous and non cencerous cells. The phosphorylation of C-Src family of proteins are linked to cancer progression by promoting, dysregulated growth survival, invasion, and migration (Sen and Johnson, 2011).

*Serine/Threonine-protein Kinase (Chk-2)*

This threonine/serine-protein kinase, which is a versatile effector that arrest checkpoint-mediated cell cycle, is involved in DNA repair and apoptosis when cell undergoes uncontrolled growth (Zannini, Delia, and Buscemi, 2014; Xu, Tsvetkov, and Stern, 2002). Chk-2 resides as non-active monomeric form in the nucleus during normal cell growth. However, ATM phosphorylates T68 and other residues of Chk-2 following DNA damage; this phosphorylation changes the conformation of Chk-2 kinase with subsequent dimerization. Dimerization promotes autophosphorylation of T383, T387, S516, and S260, T432 residues of its T-loop and kinase domain, respectively (Wu and Chen, 2003). Additional conformational changes dissociate the Chk-2 dimers into fully active monomers. Once Chk-2 is activated, it functions to phosphorylate numerous effectors including some phosphatases like CDC25A, CDC25B, and CDC25C. Phosphorylation of phosphatases by Chk-2 kinase inhibits their activity. Most importantly, phosphorylated CDC25 phosphatase blocks cell cycle progression by inhibition of CDK-cyclin complex tyrosine phosphorylation. In this way, it regulates cell cycle checkpoint arrest (Sur and Agrawal, 2016). Although in human melanoma P53 mutations are rare, melanoma cells still avoid apoptosis by over expressing a p53 negative regulator MDM4, that binds with p53 and halts its tumor suppressor activity. Approximately 65% of MDM4 were found to be upregulated in stage I-IV human melanomas. Chk-2 kinase phosphorylates MDM4, thereby reducing the degradation of p53, and controls the transcription of pro-apoptotic genes through phosphorylation of the transcription factor E2F1 (Gembarska *et al.*, 2012). Furthermore, Chk-2 kinase phosphorylates Ser-20 p53, thereby disrupting the binding between p53 and its other negative regulator MDM2, and leading to the accumulation of free active p53 in cell cytosol. Later on, p53 translocates into the nucleus, and transcribes genes involved in the apoptosis (Zannini, Delia, and Buscemi, 2014).

**Other Molecules under Investigation in Melanoma Pathogenesis**

Together with these oncogenic pathways, there are several other molecules including P27, c-KIT, melanoma tumor antigen p97 (melanotransferrin; MFI2/MTf), p53 and HSP that also play crucial roles in melanoma progression.

*Tumor Protein p53*

The tumor suppressor transcription factor p53 is a master regulator of cell growth. During different cellular stress situations, p53 either accelerates the DNA repair by inhibiting cell cycle progression or in case of damage, unrepaired cells initiate apoptosis by up-regulating the respective genes (Farnebo, Bykov, and Wiman, 2010). P53 stabilization and function has been known to be regulated by phosphorylation *via* intracellular serine and threonine kinases at many different serine, and threonine residues, including serine 15, 46, and 392 residing at *N*- and *C*-terminal regions of the protein. Phosphorylation at the CK2 site S392 of the p53 is take place following UV irradiation (Ashcroft, and Vousden, 1999). Phosphorylation of S46 of the p53, on the other hand, is proposed to regulate the pathways in response to DNA damage. S46 Phosphorylation site has been known to be regulated by several kinases, such as p38 mitogen-activated protein kinase (p38 MAPK) or AMP-activated protein kinase catalytic subunit α (AMPKα). Thus, for the regulation of normal cell function, phosphorylation of p53 might be a vital modification (Smeenk *et al.*, 2011).

*Heat shock proteins (HSP)*

In response to stress conditions like exposure to cold, temperature, UV light, oxidative stress, wound healing, tissue remodeling, cells express stress specific proteins, termed heat shock proteins (Lianos *et al.*, 2015). Heat shock proteins are also expressed during certain diseases such as gastric cancers, hepatocellular carcinoma, breast cancers, lung cancers, and colonic tumors. This kind of proteins often shows chaperone function, as they aid to correct the folding of protein that are previously damaged by the stress conditions, thereby helping them to be stabilized (Tkacova and Angelovicova, 2012).

*Heat shock proteins 27 (HSP 27)*

Small molecular chaperone HSP27 is found as large oligomers in non-stressed cells. During stress phosphorylation of HSP27 dissociates HSP27 oligomers into active dimers that provide cell cytoprotection by modulating reactive oxygen species *via* raising glutathione levels (Belhadj Slimen *et al*., 2014). Phosphorylated HSP27 is also found to oppose TNF-α-induced apoptosis by activation of transforming growth factor factor β-activated kinase 1 (TAK1) and TAK1-p38/ER K pro-survival signaling (Katsogiannou, Andrieu, and Rocchi, 2014). Phosphorylated HSP27 has found to induce invasion, enhance cell proliferation, and suppresses Fas-induced apoptosis in human PCa cells. Dephosphorylation of HSP27 has been shown its significant role in actin reorganization, and inhibition of cell adhesion (Lee *et al*., 2008). Therefore, the approach of HSP27 dephosphorylation has been used for the development of HSP-targeting/ HSP27 dephosphorylating agents for the treatment of many pathological conditions, such as oncogenesis, neurodegeneration, and senescence (Hajer *et al.,* 2017).

**Melanoma Signaling Pathways**
*MAPK Signaling Pathway*

Mitogen activated protein kinase (MAPK) signal transduction pathway comprises kinases such as RAS, RAF, MEK, and ERK These kinases are activated when receptor tyrosine kinases (RTKs) bind to their ligands on the cell membrane (Fig. 1. 1). ERK signaling is also stimulated by adhesion signals *via* integrins (Paluncic *et al*., 2016). Signals are transmitted to inactive membrane localized GDP-bound RAS, convert it into an active GTP-bound RAS. The activated form of RAS that can potentiate cell survival by activating PI3K-AKT signaling, or it can phosphorylate and activate RAF (RAF-1, BRAF, and ARAF). The activated RAF phosphorylates ERK1/2, and dissociates it from the MAPK complex. The activated ERK1/2 then translocates it into the nucleus with increased expression of an arry of genes of nuclear transcription factors, such as c-MYC and cyclin D1 leading to cell proliferation, differentiation, and survival (Inamdar *et al*., 2010). In melanocytes, MAPK pathway is tightly regulated. However, in melanomas, due to the BRAF and RAS mutation or other genetic or epigenetic modifications, the MAPK pathway is hyper phyphorylated and activated. This leads to the promotion of uncontrolled development of melanoma**,** cell survival and cell proliferation. Therefore, the MAPK pathway has been the focus of intense investigation in the field of oncology as a potentially important therapeutic target (Amaral *et al*., 2017).

*PI3K/AKT /PTEN Signaling Pathway*

This pathway is involved in melanoma proliferation, and metabolism. The kinases of this pathway are highly conserved phosphatidylinositol-4,5-bisphosphate 3-kinase (PI3K).(Paluncic *et al*., 2016). Binding of insulin-like growth factor-1 (IGF-1) receptor with its ligand, insulin-like growth factor-1, results in the phosphorylation of phosphatidylinositol-4,5-bisphosphate (PIP2), to as a binding place for 3-phosphoinositide-dependent protein kinase 1(PDK1) which activates the protein kinase B along with several other kinases (Siroy, Davies, and Lazar, 2016). Protein kinase B, also known as AKT, and a crucial effector protein, has three isoforms (AKT1, AKT2, and AKT3). AKT3 is frequently activated in melanoma. AKT3 activation may result in phosphorylation of up to 100 target proteins (Risso *et al*., 2015), including p70 ribosomal S6 protein kinase (S6K), glucocorticoid-dependent kinase (SGK), p90 ribosomal protein S6 kinase (RSK), and atypical PKC isoforms. Glycogen synthase kinase 3 beta (GSK3β) is one of the substrates of p-AKT that loses its kinase activity after phosphorylation by p-AKT. One of the phosphorylation targets of GSK3β is β-catenin that is not phosphorylated when GSK3β has lost its kinase activity. Hence, unphosphorylated β-catenin translocates and accumulates in the nucleus and subsequently up-regulates the expression of c-MYC and cyclin D1 oncogenic genes. In addition, AKT can be completely phosphorylated, and activated by the mammalian target of rapamycin (mTOR). The phosphorylated AKT can also phosphorylate and activate ERK with a consequent increased in the expression of c-MYC. The phosphorylated AKT mediates anti-apoptotic effect by these regulatory mechanisms, leading to cancer progression (Manning and Toker, 2017). However, a component of MAPK signaling, *i.e.,* oncogenic, small GTPase RAS, is also known to be a positive upstream regulator of the PI3K pathway. Active PI3K activates two diverse but interrelated pathways: the RAF–MEK–ERK pathway, and the PI3K-AKT signaling pathway. The latter provides activation of AKT, and its downstream targets. The PI3K-AKT pathway also stimulates anabolism, while the RAF–MEK–ERK pathway is more important in proliferation, and invasion (Gomes and Blenis, 2015).

*The JAK-STAT Signaling Pathway*

Signal transducer and activator of transcription (STAT) forms a latent intracellular transcription factor family of proteins. The STATs family of protein includes STAT1, STAT2, STAT3, STAT4, STAT5α, STAT5β, and STAT6. These transcription factors reside within the cytosol in monomeric, non-phosphorylated form (O'Shea *et al*., 2015). Non-phosphorylated STATs become activated in response to stimulated cells with growth factors or cytokines receptors after binding with their cognate ligands. This binding results in dimerization of these receptors, and activation of the intrinsic receptor tyrosine kinase, together with receptor-associated tyrosine kinase JAKs, and Src-activated STAT. Furthermore, these kinases can subsequently phosphorylate the receptor cytosolic tail. This phosphorylation creates a docking site where monomeric non-phosphorylated STAT and Src recruits them *via* their SH2 domain, and thereby themselves become substrates for tyrosine phosphorylation (Aittomäki, and Pesu, 2014). In addition to the activation of STAT *via* receptor-associated tyrosine kinases (JAKs and Src) by both growth factors, and cytokine receptors, Src and Abl phosphorylate STATs in a receptor independent reaction (Berger *et al.,* 2014). Phosphorylation of STAT monomers induces dimerization, and subsequent translocation to the nucleus, where gene expression after binding to STAT DNA-response elements. By this mechanism the STATs family of protein are supposed to up-regulate the apoptosis inhibitors genes (Bcl-xL, Mcl-1), cell cycle regulators (cyclins D1/D2, c-Myc), and inducers of angiogenesis (VEGF). They are also considered to participate in oncogenesis through their antiapoptotic effects, as well as proliferation, and angiogenesis (Miklossy, Hilliard, and Turkson, 2013).

*MSK Signal Transduction Pathway*

The mitogen- and stress-activated kinases-1 and -2 (MSK1, and MSK2) serine/ threonine kinases are required to activate bZIP cellular transcription factor cAMP response element-binding protein (CREB) by phosphorylating it on Ser133 residue. Numerous genes of cell proliferation, cell survival, memory, glucose homeostasis, and spermatogenesis are transcribed when CREB binds to their promoter sequences (Kar *et al*., 2017). However, MSK1/2 is itself activated after phosphorylation by two mitogen-activated protein kinases (MAPKs). MAPKs are intracellular transducers of signals for mediating signals from numerous physiological stimuli, resulting in the regulation of a broad array of cellular responses. These ubiquitous and major MAPK families regulate a plethora of physiological processes by controlling the function of hundreds of cellular proteins *via* phosphorylation, either directly or *via* other kinases such as MSK (Duda and Frödin, 2013).

*Mammalian Target of Rapamycin (mTOR) Signal Transduction Pathway*

"Mechanistic Target of Rapamycin" (mTOR) is a serine/ threonine member of the phosphatidylinositol 3-kinase-related kinase family. It is an enzymatic complex of two distinct proteins, *i.e.,* mTOR complex 1 (mTORC1), and mTOR complex 2 (mTORC2). Due to the core component of two distinct complexes, mTOR exhibits dual actions. mTOR functions as a cell growth regulator in the form of mTORC1, and as a regulator of cell survival in the form of mTORC2. (Gaubitz, Prouteau, Kusmider, and Loewith, 2016). The activity of mTORC1 is regulated under the influence of various inputs, including certain mechanical stimuli, insulin, rapamycin, growth factors, phosphatidic acid, certain amino acids, and their derivatives (*e.g*., L-leucine and β-hydroxy-β-methylbutyric acid), and oxidative stress, resulting in transcription of genes responsible for protein synthesis. Eventually, by controlling protein synthesis it regulates cell proliferation, and cell survival. Altogether, mTORC1 functions as a nutrient/energy/redox sensor. Moreover, mTORC1. also regulates the transcription of genes of lipid synthesis. It also has a role in cell motility, and autophagy (Kim, Buel, and Blenis, 2013).

Together with various stimuli, the function of mTORC1 is regulated by its negative regulator p70 ribosomal protein S6 kinase (p70S6K) that is itself a serine/threonine kinase. Normally, this kinase stays bound to mTORC1, and halts its activity. However, in response to cell proliferation and survival stimuli, its PRAS40 becomes phosphorylated, and dissociates from mTORC1, allowing it to function in protein, ribosomes, nucleotide, lipid synthesis, and in promoting cell growth, and blocking autophagy (Völkers and Sussman, 2013).

mTORC2 is known to regulate AKT kinase by phosphorylating serine 473 of AKT/PKB. Phosphorylation of serine 473 by mTORC2 induces AKT phosphorylation on threonine 308 by PDK1, and leads to complete AKT activation. Fully activated AKT regulates cell survival, and metabolism (Rad, Murray, and Tee, 2018). In tumor and normal cells, mTOR2 maintains actin organization by activating paxillin, F-actin stress fibers, Rac1, RhoA, Cdc42, and protein kinase Cα. Consequently, its dysregulation is associated with cancerous cell migration and invasion (Holroyd, and Michie, 2018). Both mTORC1, and mTORC2 are regulated by phosphorylation, complex formation, and localization. The mTOR pathway is regarded as the central regulator of mammalian physiology and metabolism with important roles in the functioning of diverse organs, in various cancers, and human diseases where it was found to be dysregulated (Giguère, 2018).

*G Protein-coupled receptors (GPCRs) Signaling Pathways WNT Signaling Pathway*

This pathway comprises various members of secreted lipid-modified signaling glycoproteins that orchestrate and influence a myriad of biological, and developmental processes in the cell. Hyperactivation of the WNT pathway leads to the progression of various types of cancers, including melanoma (Willert and Nusse, 2012). WNT is a glycoprotein that is the cognate lingand of FZD receptors. The binding of WNT with FZD receptors initiates three possible pathways including (i) a classical intracellular signal transduction pathway that depends on β-catenin, (ii) a non- classical signal transduction pathway that although not dependent on β-catenin, however, controls cell polarity, and (iii) a WNT, and protein Kinase-C (PKC)-dependent pathway. Both classical and non- classical WNT signaling pathways regulate tumor progression at different stages. Whereas the non-classical pathway mainly regulates cancer metastasis, while the classical pathway mainly works to transform melanocytes into melanoma (Paluncic *et al*., 2016).

*The Classical WNT Pathway*

The mechanism of this pathway is based on transcription factor β-catenin regulation *via* activation, and degradation. In the absence of WNT signaling, axin interacts with β-catenin adenomatous polyposis coli (APC), glycogen synthase kinase 3β (GSK3β), and casein kinase 1α (CK1α) with separate domains to make a complex. These complex is able to coordinate a sequential phosphorylation of β-catenin, CKI, and GSK3, thus targeting them for ubiquitination, and subsequent degradation by the proteasome (Paluncic *et al.*, 2016). However, when WNT, together with its co-receptor the low-density-lipoprotein-related proteins 5/6 (LRP5/6), binds with the FZD receptor on cell membrane, β-catenin is released from the complex, and stabilizes it in the cytosol. From the cytosol it is subsequently translocated into the nucleus, where it transcribes WNT-responsive genes, including cyclin D1, and c-MYC, and ZEB-1 (Kaur, Webster and Weeraratna, 2016).

*ROS Mediated Pathways of Cell Migration and Metastasis and Proliferation*

Not only activation of kinases *via* phosphorylation can result in cell transformation, survival, and proliferation, but oxidants have also been shown to trigger the phosphorylation cascades of MAPKs and PI3K/ATK pathways. Three MAPKs pathways are extracellular signal-regulated kinase (ERK), c-Jun *N*-terminal kinase (JNK), and p38 MAPK. They are structurally related but functionally distinct kinases, considered to be modulated by cellular redox balance. In addition to cytokine and growth factor receptors activation, their activation results in uncontrolled cell proliferation, and survival. ERKs are activated by mitogens, while JNK and p38 MAPK can be activated by heat shock proteins, inflammatory cytokines, and ROS (Zhao, Wang, and Tony 2015). These pathways are activated by different threshold of ROS. A high level of ROS activates JNK and NF-κB pathways while p38 MAPK-related known to induce EMT including epidermal growth factor receptor (EGFR). Activation of NF-κB and AP-1 and c-Jun transcription factors induce the production of proteins, involved in cell proliferation, differentiation, inflammation, inhibition of apoptosis, and transformation (Son *et al.*, 2011).

**Melanoma Metastasis**

90% of all cancer deaths can be attributed to metastasis, and not to the primary tumors. In this process, cancerous cells attack adjacent tissues, and spread to distant organs (Steeg, 2016). Cancer metastasis is based on highly complex molecular, and cell biological principles, collectively referred to as 'metastatic factors (Leber and Efferth, 2009). Following steps are involved in the development of a secondary tumor from the primary tumor.

*Epithelial–Mesenchymal Transition*
The biological process of epithelial–mesenchymal transition (EMT) is the initial step of developing metastatic potential. This event alters the adhesive properties between melanocytes and keratinocytes, with subsequent morphological changes from an epithelial to a mesenchymal phenotype, cell motility, and loss of polarity, that permit tumor cells to outflow from their micro environment, and to migrate to distant locations *via* circulatory blood or lymphatic systems where in a visceral organ, they create a metastasis (Mittal, 2018). The process of EMT is characterized by downregulation of several calcium-dependent adhesion molecules. Most prominently these molecules are E cadherins, claudins, occludin or catenin, and cytokeratins proteins.

However, several mesenchymal proteins have been observed to be upregulated during EMT process, which confers the ability to detach from extracellular matrix to a melanoma cell, and migrate out of its nascent environment. The process of melanoma development lies on two basic early steps: disruption of adhesive interaction between melanocytes, and keratinocytes during EMT, and the invasion and migration of melanoma cells. These processes are regulated by N and E cadherins. In addition to transforming growth factor receptor signaling pathways, there are also numerous receptor tyrosine kinase signal transductions pathways that are known to induce EMT including epidermal growth factor receptor (EGFR).

*Cell Migration and Invasion*
Cell migration is the process of cell movement from one place to the other in response to chemical signals. Cells are packed together with the help of adhesive proteins at different adhesion sites (Khalili and Ahmad, 2015). These adhesion sites include focal complexes and focal adhesion complexes. Focal adhesion complexes are integrin-containing, multi-protein complex at cell adhesion sites (Burridge and Guilluy, 2016). These proteins create links between intracellular actin, and the extracellular substrate. To initiate migration of cells the activity of these proteins need to be altered that results in modification of the cell shape and stiffness, and initiate migration (Khalili and Ahmad, 2015). Cells migrate in both normal and pathological processes due to the coordinated and dynamic regulation of these adhesions sites under the influence of signal transduction pathways. Signals for cell migration, proliferation, and apoptosis are generated outside the cells when integrin clusters at the cell surface, and growth factor receptor binds to cognate growth factors (Mitra and Schlaepfer, 2006; Chen *et al.*, 2018; Marlowe *et al.*, 2016). These signals then activate a non-receptor, non-membrane tyrosine kinase of the focal complex termed focal adhesion kinase (FAK) by phosphorylation of its autophosphorylation site (Tyrosine 397 residue). After phosphorylation, tyrosine 397 focal adhesion kinase generates a high-affinity docking site for the SH2 domain of Src family kinases. Consequently, it interacts with Src and makes an activated Src-FAK complex. The activated Src-FAK, in addition to the phosphorylation of multiple tyrosine residues Y576, Y577, Y925 of FAK, also phosphorylates other signaling molecules, such as p130Cas and paxillin. This cascade of phosphorylations results in cell cytoskeletal changes, and activation of other signaling pathways of cell migration, invasion, and metastasis (Zhao and Guan, 2011; Mitra, Hanson and Schlaepfer, 2005). The Y397 site is one of the main phosphorylation sites that regulate a multitude of cellular processes involved in cancer metastases through FAK signaling in the cells. Therefore, FAK is considered as the master regulator protein due to its ability to mediate primarily cell migration, in addition to cell survival, apoptosis, and proliferation (Tai, Chen, and Shen, 2015). Thus, inhibiting the activity of FAK by dephosphorylating its Y397 residue represents an excellent approach for treating cancer metastasis (Golubovskaya *et al.*, 2008). Besides focal adhesion kinase, other mitogen-activated protein kinases (MAPKs), including Jun *N*-terminus kinase (JNK), p38, and ERK, are also considered as essential pathways of cell migration (Sun *et al.*, 2015).

*Intravasation*
The penetration of endothelial cells by cancerous cells is under the effect of proteases that degrade the endothelium, and permit the cells to penetrate into the blood vessels

*Circulation*
Migration of cancer cell in lymphatic or blood vessel to a distant tissue takes place when cancer cells acquire anchorage independent proliferation.

*Extravasation*
In this step of metastasis, tumor cells migrate out of vessels into organs. An adhesion molecule of the immunoglobulin superfamily CD146 melanoma cell adhesion molecule MCAM is present at both endothelial junctions and tumor-endothelial contact sites, where it regulates the melanoma cell extravasation.

*Colonization*
To settle at a new location away from the primary tumor, cancer cells use diverse steps and overcome several hurdles, including infiltration into distant tissues, evading immune defenses, adapting to supportive niches, surviving as latent tumor-initiating seeds, eventually breaking out to replace the host tissue. All this makes metastasis a highly ineffective process.

**Metastatic Melanoma Treatment Options**

Systemic therapies are provided by medications that spread in the body to treat cancer cells at all possible locations. These therapies are used in stage IV melanoma patients. They include cytotoxic chemotherapy, targeted drugs, immunotherapy, hormonal therapy, and bio-chemotherapy (Bhatia, Tykodi, and Thompson, 2009).

*Metastatic Melanoma Cytotoxic Chemotherapy*
Cytotoxic chemotherapy has been used since last three decades for the treatment of metastatic melanoma, and aims at preventing or relieving symptoms. Few chemotherapeutic agents with modest antitumor efficacy have been used for the treatment of metastatic melanoma alone or in combination (Bhatia, Tykodi, and Thompson, 2009).

*Single Agent Cytotoxic Chemotherapy*
Dacarbazine is an alkylating agent, and the only standard single chemotherapeutic drug for metastatic melanoma treatment with a significant survival benefit. This drug represses gene transcription by methylating guanine N-7 or O-6 positions. Although there are some other chemotherapeutic drugs that have been approved by US-FDA for metastatic melanoma, no other therapy has yet shown a similar or greater efficacy over dacarbazine (Hill, Krementz, and Hill, 1984).

Despite some common side effects like mild nausea and vomiting, myelosuppression, and fatigue, most patients are able to maintain their basic quality of life with dacarbazine (Bhatia, Tykodi, and Thompson, 2009).

Temozolomide is an analog of dacarbazine. It has the same mechanism of action like dacarbazine, and has been investigated extensively in metastatic melanoma due to its excellent oral bioavailability, and ability for CNS penetration (Patel *et al.*, 2011; Chiarion-Sileni, 2008).

Anti-microtubular agents, *i.e., Vinca* alkaloids (vincristine, vinblastine), and taxanes (paclitaxel, and nab-paclitaxel) are anti-microtubule agents. They exhibit modest activity in patients with metastatic melanoma. These agents used sometimes in combination with cytotoxic chemotherapy or biochemotherapy regimens (Nathan Berd, and Mastrangelo, 2000).

Platinum analogs interfere with DNA replication. In metastatic melanoma, platinum drugs cisplatin and carboplatin have limited activity (Evans, Casper, and Rosenbluth, 1987).

Nitrosoureas are comparable to dacarbazine with more myelosuppression and alopecia. This class of drug includes carmustine (BCNU), lomustine (CCNU), and fotemustine, that can be used as single agents. Patients with brain metastases are effectively treated with fotemustine (Khayat *et al.*, 1994).

*Combination Chemotherapy*
The modest anti-tumor activity of the chemotherapeutic agents, mentioned in the preceding sections has led to the investigation of their combinations. Studies have been conducted to investigate chemotherapy in terms of response rate, and possibly survival.

The Dartmouth regimen is combination of four drugs, cisplatin, dacarbazine, carmustine, and tamoxifen (Del Prete *et al.*, 1984). However, Dartmouth regimen did not show a statistically significant benefit in favor of the combination *vs.* dacarbazine monotherapy in a phase III randomized clinical trial (Chapman *et al.*, 1999).

CVD regimen, is a regmen that includes vinblastine, cisplatin, and dacarbazine and used for the treatment of advanced-stage melanoma (Legha *et al.*, 1989; Legha *et al.*, 1996).

Paclitaxel and carboplatin (PC) regimen in both metastatic melanoma patients, and patients who have received prior chemotherapy, has been used as a combination therapy (Hodi *et al.*, 2002).

Pro-apoptotic drugs regimen is a combination of cisplatin, with 5-fluorouracil (5-FU). This is the recent major form of proapoptotic drug treatment, used in the clinical management

*Metastatic Melanoma Immunotherapy*
It includes interleukin-2 and interferon alfa-2b for treating metastatic melanoma by enhancing immune response against the tumor cells (Martin and Lo, 2018; Atkins *et al.,* 1999; Creagan *et al.,* 1986)

*Metastatic Melanoma Targeted Therapies*
These therapies include BRAF-inhibitors vemurafenib that interferes with the B-Raf/MEK/ERK pathway, dabrafenib that inhibits some malformed BRAF kinases, and trametinib for inhibiting MEK-1/2.

*Metastatic Melanoma Bio-Chemotherapy*
Biochemotherapy combines the use of cytotoxic agents with IFN-α and/or IL-2. Although biochemotherapy compared to chemotherapy alone provides increased response rate in patients who have rapidly progressing disease, due to toxicities associated with biochemotherapy, together with its insignificant survival alternatives, such as HD IL-2 or other agents, are used for patients who suffer from chronic melanoma (Ives *et al*., 2007: Lui *et al*., 2007).

*Novel Therapies under Investigation*
Currently, anti-CTLA-4 anti-body, adoptive cell therapy, and vaccination are being used as novel approaches. Certain vaccines, based on peptides (e.g., gp100), nucleic acids (e.g., Allovectin-7), dendritic cells, and heat shock protein complexes (e.g., vitespen), have given encouraging results in phase II and phase III clinical trials (Richards *et al*., 2005; Dudley *et al*., 2008). Even with these available treatment options, metastatic melanoma still presents itself as an incurable disease.

## Conclusion

In order to obtain a successful treatment of aggressive cancer, it is important to develop, and use new approaches for metastatic melanoma prevention, and treatment in cases of refractory treatments. Various new therapeutic agents have been developed against various protein kinases, that are considered as molecular players of the currently investigated pathways described above. They are known to be involved in metastatic melanoma. Some of the new therapeutic interventions against these kinases have successfully reached clinical trials (Harris *et al*., 2016) while others already have been approved by the US-FDA. However, due to the emergence of resistance against these drugs, and the adverse effects they cause, multiple phases II and III melanoma trials are currently underway that are either studying the effect of combination treatments, or new synthetic and natural compounds as new possible treatment options. However, no systemic therapy has meaningfully changed these survival end points so far, and there is still a need to prevent the induction of melanoma or develop combination therapies that target the unique molecular profile of melanoma tumors. Additionally, there is also need for identification of new synthetic, and natural compounds as new treatment options.

## Conflicts of Interest

## Authors' Contributions

## Acknowledgments

The authors are thankful to the International Union of Biochemistry and Molecular Biology (IUBMB) for two consecutive grants to Q-u-A, the Hussain Ebharim Jamal Research Institute of Chemistry (H.E.J) for a six months Prof. Saleem- uz- Zaman grant to Q-u-A and Dr. Panjwani Center for Molecular Medicine and Drug Research for a five year PCMD grant to Q-u-A.

## References

Addison, C. L. (**2014**). Epidermal growth factor receptor. *Encyclopedia of Cancer*, 1-7.

Aittomäki, S., and Pesu, M**.** (**2014**). Therapeutic targeting of the JAK/STAT pathway. *Basic and Clinical Pharmacology and Toxicology*, *114*(1), 18-23.

Amaral, T., Sinnberg, T., Meier, F., Krepler, C., Levesque, M., Niessner, H., and Garbe, C. (**2017**). The mitogen-activated protein kinase pathway in melanoma part I - Activation and primary resistance mechanisms to BRAF inhibition. *European Journal of Cancer*, 73, 85-92.

Angus, S. P., Zawistowski, J. S., and Johnson, G. L. (**2018**). Epigenetic mechanisms regulating adaptive responses to targeted kinase inhibitors in cancer. *Annual Review of Pharmacology and Toxicology*, *58*, 209-229.

Ashcroft, M., Kubbutat, M. H., and Vousden, K. H. (**1999**). Regulation of p53 function and stability by phosphorylation. *Molecular and Cellular Biology*, *19*(3), 1751-1758.

Atkins, M. B., Lotze, M. T., Dutcher, J. P., Fisher, R. I., Weiss, G., Margolin, K., and Paradise, C. (**1999**). High-dose recombinant interleukin 2 therapy for patients with metastatic melanoma: Analysis of 270 patients treated between 1985 and 1993. *Journal of Clinical Oncology*, *17*(7), 2105-2105.

Bandarchi, B., Jabbari, C. A., Vedadi, A., and Navab, R. (**2013**). Molecular biology of normal melanocytes and melanoma cells. *Journal of Clinical Pathology*, *66*(8), 644-648.

Bandarchi, B., Ma, L., Navab, R., Seth, A., and Rasty, G. (**2010**). From melanocyte to metastatic malignant melanoma. *Dermatology Research and Practice*, 2010.

Belhadj Slimen, I., Najar, T., Ghram, A., Dabbebi, H., Ben Mrad, M., and Abdrabbah, M. (**2014**). Reactive oxygen species, heat stress and oxidative-induced mitochondrial damage. A review. *International Journal of Hyperthermia*, *30*(7), 513-523.

Berger, A., Hoelbl-Kovacic, A., Bourgeais, J., Hoefling, L., Warsch, W., Grundschober, E., and Schuster, C. (**2014**). PAK-dependent STAT5 serine phosphorylation is required for BCR-ABL-induced leukemogenesis. *Leukemia*, *28*(3), 629.

Berridge, M. J. (**2014**). Module 1: Introduction. *Cell Signalling Biology*, 10-14.

Bhatia, S., Tykodi, S. S., and Thompson, J. A. (**2009**). Treatment of metastatic melanoma: An overview. *Oncology*, *23*(6), 488.

Bisevac, J. P., Djukic, M., Stanojevic, I., Stevanovic, I., Mijuskovic, Z., Djuric, A., and Vojvodic, D. (**2018**). Association between oxidative stress and melanoma progression. *Journal of Medical Biochemistry*, *37*(1), 12-20.

Bononi, A., Agnoletto, C., De Marchi, E., Marchi, S., Patergnani, S., Bonora, M., and Pinton, P. (**2011**). Protein kinases and phosphatases in the control of cell fate. *Enzyme Research*, 1-26.

Burridge, K., and Guilluy, C. (**2016**). Focal adhesions, stress fibers and mechanical tension. *Experimental Cell Research*, *343*(1), 14-20.

Cain, J. A., Solis, N., and Cordwell, S. J. (**2014**). Beyond gene expression: The impact of protein post-translational modifications in bacteria. *Journal of Proteomics*, *97*, 265-286.

Chapman, P. B., Einhorn, L. H., Meyers, M. L., Saxman, S., Destro, A. N., Panageas, K. S., and Houghton, A. N. (**1999**). Phase III multicenter randomized trial of the Dartmouth regimen versus dacarbazine in patients with metastatic melanoma. *Journal of Clinical Oncology*, *17*(9), 2745-2745.

Chen, Z., Oh, D., Dubey, A. K., Yao, M., Yang, B., Groves, J. T., and Sheetz, M. (**2018**). EGFR family and Src family kinase interactions: Mechanics matters?. *Current Opinion in Cell Biology*, 51, 97-102.

Cheng, H. C., Qi, R. Z., Paudel, H., and Zhu, H. J. (**2011**). Regulation and function of protein kinases and phosphatases. *Enzyme Research*, 2011, 1-3.

Chiarion-Sileni, V., Guida, M., Ridolfi, R., Romanini, A., Brugnara, S., Del Bianco, P., and De Salvo, G. (**2008**). Temozolomide (TMZ) as prophylaxis for melanoma brain metastases (BrM): Results from a phase III, multicenter study. *Journal of Clinical Oncology*, *26*(15), 20014-20014.

Creagan, E. T., Ahmann, D. L., Frytak, S., Long, H. J., Chang, M. N., and Itri, L. M. (**1986**). Phase II trials of recombinant leukocyte A interferon in disseminated malignant melanoma: Results in 96 patients. *Cancer Treatment Reports*, *70*(5), 619-624.

Day, E. K., Sosale, N. G., and Lazzara, M. J. (**2016**). Cell signaling regulation by protein phosphorylation: a multivariate, heterogeneous, and context-dependent process. *Current Opinion in Biotechnology*, *40*, 185-192.

Dayem, A. A., Choi, H. Y., Kim, J. H., and Cho, S. G. (**2010**). Role of oxidative stress in stem, cancer, and cancer stem

Del Prete, S. A. (**1984**). Combination chemotherapy with cisplatin, carmustine, dacarbazine and tamoxifen in metastatic melanoma. *Cancer Treatment Review*, 68, 1403-1405.

Duda, K., and Frödin, M. (**2013**). Stimuli that activate MSK in cells and the molecular mechanism of activation. *Madame Curie Bioscience Database*, 1-46.

Dudley, M. E., Yang, J. C., Sherry, R., Hughes, M. S., Royal, R., Kammula, U., and Wunderlich, J. (**2008**). Adoptive cell therapy for patients with metastatic melanoma: Evaluation of intensive myeloablative chemoradiation preparative regimens. *Journal of Clinical Oncology*, *26*(32), 5233.

Espada, J., and Martin-Perez, J. (**2017**). An update on Src family of nonreceptor tyrosine kinases biology. *International Review of Cell and Molecular Biology,* 331, 83-122.

Evans, L. M., Casper, E. S., and Rosenbluth, R. (**1987**). Phase II trial of carboplatin in advanced malignant melanoma. *Cancer Treatment Reports*, *71*(2), 171-172.

Farnebo, M., Bykov, V. J., and Wiman, K. G. (**2010**). The p53 tumor suppressor: a master regulator of diverse cellular processes and therapeutic target in cancer. *Biochemical and Biophysical Research Communications*, *396*(1), 85-89.

Gaubitz, C., Prouteau, M., Kusmider, B., and Loewith, R. (**2016**). TORC2 structure and function. *Trends in Biochemical Sciences*, *41*(6), 532-545.

Gembarska, A., Luciani, F., Fedele, C., Russell, E. A., Dewaele, M., Villar, S., and Goydos, J. (**2012**). MDM4 is a key therapeutic target in cutaneous melanoma. *Nature Medicine*, 18(8), 1239.

Giguère, V. (**2018**). Canonical signaling and nuclear activity of mTOR: a teamwork effort to regulate metabolism and cell growth. *The FEBS Journal*. 285(2018), 1572-1588.

Golubovskaya, V. M., Nyberg, C., Zheng, M., Kweh, F., Magis, A., Ostrov, D., and Cance, W. G. (**2008**). A small molecule inhibitor, 1, 2, 4, 5-benzenetetraamine tetrahydrochloride, targeting the y397 site of focal adhesion kinase decreases tumor growth. *Journal of Medicinal Chemistry*, *51*(23), 7405-7416.

Gomes, A. P., and Blenis, J. (**2015**). A nexus for cellular homeostasis: the interplay between metabolic and signal transduction pathways. *Current Opinion in Biotechnology*, 34, 110-117.

Grangeasse, C., Stülke, J., and Mijakovic, I. (**2015**). Regulatory potential of post-translational modifications in bacteria. *Frontiers in Microbiology*, 6, 500.

Hajer, Z., Claudia, A., Erik, L., Sara, K., Maurizio, F., Ridha, O., and Maria, K. (**2017**). Targeting Hsp27/eIF4E interaction with phenazine compound: A promising alternative for castration-resistant prostate cancer treatment. *Oncotarget*, *8*(44), 77317.

Harris, Z., Donovan, M. G., Branco, G. M., Limesand, K. H., and Burd, R. (**2016**). Quercetin as an emerging anti-melanoma agent: A four-focus area therapeutic development strategy. *Frontiers in Nutrition*, 3, 48.

Hill, G. J., Krementz, E. T., and Hill, H. Z. (**1984**). Dimethyl.

Hill, G. J., Krementz, E. T., and Hill, H. Z. (**1984**). Dimethyl triazeno imidazole carboxamide and combination therapy for melanoma IV. Late results after complete response to chemotherapy (Central Oncology Group protocols 7130, 7131, and 7131A). *Cancer*, *53*(6), 1299-1305.

Ho, A., Jönsson, G., and Tsao, H. (**2017**). Melanoma Genetics and Genomics. *Melanoma Development*, 63-93.

Hodi, F. S., Soiffer, R. J., Clark, J., Finkelstein, D. M., and Haluska, F. G. (**2002**). Phase II study of paclitaxel and carboplatin for malignant melanoma. *American Journal of Clinical Oncology*, *25*(3), 283-286.

Holroyd, A. K., and Michie, A. M. (**2018**). The role of mTOR-mediated signaling in the regulation of cellular migration. *Immunology Letters*.

Inamdar, G. S., Madhunapantula, S. V., and Robertson, G. P. (**2010**). Targeting the MAPK pathway in melanoma: Why some approaches succeed and other fail. *Biochemical Pharmacology*, *80*(5), 624-637.

Ives, N. J., Stowe, R. L., Lorigan, P., and Wheatley, K. (**2007**). Chemotherapy compared with biochemotherapy for the treatment of metastatic melanoma: A meta-analysis of 18 trials involving 2,621 patients. *Journal of Clinical Oncology*, *25*(34), 5426-5434.

Kar, R. K., Kharerin, H., Padinhateeri, R., Bhat, P. J., and Yao, J. (**2017**). Cyclic AMP-response element-binding protein (CREB) plays key transcriptional roles in cell metabolism, proliferation, and survival. *Journal of Molecular Biology*, *429*(1), 142-157.

Katsogiannou, M., Andrieu, C., and Rocchi, P. (**2014**). Heat shock protein 27 phosphorylation state is associated with cancer progression. *Frontiers in Genetics*, 5, 346.

Kaur, A., Webster, M. R., and Weeraratna, A. T. (**2016**). In the WNT-er of life: WNT signalling in melanoma and ageing. *British Journal of Cancer*, *115*(11), 1273.

Kemp, B. E. (**2018**). Peptides and protein phosphorylation. *CRC Press*.

Khalili, A. A., and Ahmad, M. R. (**2015**). A review of cell adhesion studies for biomedical and biological applications. *International Journal of Molecular Sciences, 16*(8), 18149-18184.

Khayat, D., Giroux, B., Berille, J., Cour, V., Gerard, B., Sarkany, M., and Bizzari, J. P. (**1994**). Fotemustine in the treatment of brain primary tumors and metastases. *Cancer Investigation*, *12*(4), 414-420.

Kim, S. G., Buel, G. R., and Blenis, J. (**2013**). Nutrient regulation of the mTOR complex 1 signaling pathway. *Molecules and Cells*, *35*(6), 463-473.

Krebs, E. G., and Beavo, J. A. (**1979**). Phosphorylation-dephosphorylation of enzymes. *Annual Review of Biochemistry*, *48*(1), 923-959.

Kumar, V., Abbas, A. K., and Aster, J. C. (**2017**). Robbins Basic Pathology E-Book. Elsevier Health Sciences.

Leber, M. F., and Efferth, T. (**2009**). Molecular principles of cancer invasion and metastasis. *International Journal of Oncology*, *34*(4), 881-895.

Lee, J. W., Kwak, H. J., Lee, J. J., Kim, Y. N., Lee, J. W., Park, M. J., and Lee, J. S. (**2008**). HSP27 regulates cell adhesion and invasion *via* modulation of focal adhesion kinase and MMP-2 expression. *European Journal of Cell Biology*, *87*(6), 377-387.

Legha, S. S., Ring, S., Bedikian, A., Plager, C., Eton, O., Buzaid, A. C., and Papadopoulos, N. (**1996**). Treatment of metastatic melanoma with combined chemotherapy containing cisplatin, vinblastine and dacarbazine (CVD) and biotherapy using interleukin-2 and interferon-α. *Annals of Oncology*, *7*(8), 827-835.

Lewis-Tuffin, L. J., Feathers, R., Hari, P., Durand, N., Li, Z., Rodriguez, F. J., and Anastasiadis, P. Z. (**2015**). Src family kinases differentially influence glioma growth and motility. *Molecular Oncology*, *9*(9), 1783-1798.

Lianos, G. D., Alexiou, G. A., Mangano, A., Mangano, A., Rausei, S., Boni, L., and Roukos, D. H. (**2015**). The role of heat shock proteins in cancer. *Cancer Letters*, *360*(2), 114-118.

Lui, P., Cashin, R., Machado, M., Hemels, M., Corey-Lisle, P. K., and Einarson, T. R. (**2007**). Treatments for metastatic melanoma: Synthesis of evidence from randomized trials. *Cancer Treatment Reviews*, *33*(8), 665-680.

Manning, B. D., and Toker, A. (**2017**). AKT/PKB signaling: Navigating the network. *Cell*, *169*(3), 381-405.

Marlowe, T. A., Lenzo, F. L., Figel, S. A., Grapes, A. T., and Cance, W. G. (**2016**). Oncogenic receptor tyrosine kinases directly phosphorylate focal adhesion kinase (FAK) as a resistance mechanism to FAK-kinase inhibitors. *Molecular Cancer Therapeutics*, *15*(12), 3028-3039.

Martin, S., and Lo, R. (**2018**). Update in Immunotherapies for Melanoma. *Biologic and Systemic Agents in Dermatology*, 549-552.

Mattia, G., Puglisi, R., Ascione, B., Malorni, W., Carè, A., and Matarrese, P. (**2018**). Cell death-based treatments of melanoma: Conventional treatments and new therapeutic strategies. *Cell Death and Disease*, *9*(2), 112.

Merlot, A. M., Huang, M. L. H., and Richardson, D. R. (**2016**). Roads to melanoma: Key pathways and emerging players in melanoma progression and oncogenic signaling. *Biochimica et Biophysica Acta (BBA)-Molecular Cell Research*, 1863(4), 770-784.

Miklossy, G., Hilliard, T. S., and Turkson, J. (**2013**). Therapeutic modulators of STAT signalling for human diseases. *Nature Reviews Drug Discovery*, *12*(8), 611.

Mitra, S. K., and Schlaepfer, D. D. (**2006**). Integrin-regulated FAK–Src signaling in normal and cancer cells. *Current Opinion in Cell Biology*, *18*(5), 516-523.

Mitra, S. K., Hanson, D. A., and Schlaepfer, D. D. (**2005**). Focal adhesion kinase: In command and control of cell motility. *Nature Reviews Molecular Cell Biology*, *6*(1), 56.

Mittal, V. (**2018**). Epithelial mesenchymal transition in tumor metastasis. *Annual Review of Pathology: Mechanisms of Disease*, *13*, 395-412.

Nathan, F. E., Berd, D., Sato, T., and Mastrangelo, M. J. (**2000**). Paclitaxel and tamoxifen. *Cancer*, *88*(1), 79-87.

Nishi, H., Shaytan, A., and Panchenko, A. R. (**2014**). Physicochemical mechanisms of protein regulation by phosphorylation. *Frontiers in Genetics*, 5, 270.

O'Shea, J. J., Schwartz, D. M., Villarino, A. V., Gadina, M., McInnes, I. B., and Laurence, A. (**2015**). The JAK-STAT pathway: Impact on human disease and therapeutic intervention. *Annual Review of Medicine*, 66, 311-328.

Paluncic, J., Kovacevic, Z., Jansson, P. J., Kalinowski, D., Legha, S. S., Ring, S., Papadopoulos, N., Plager, C., Chawla, S., and Benjamin, R. (**1989**). A prospective evaluation of a triple-drug regimen containing cisplatin, vinblastine, and dacarbazine (CVD) for metastatic melanoma. *Cancer*, *64*(10), 2024-2029.

Patel, P. M., Suciu, S., Mortier, L., Kruit, W. H., Robert, C., Schadendorf, D., and Becker, J. (**2011**). Extended schedule, escalated dose temozolomide versus dacarbazine in stage IV melanoma: Final results of a randomised phase III study (EORTC 18032). *European Journal of Cancer*, *47*(10), 1476-1483.

Pearson, R. B., and Kemp, B. E. (**1991**). [3] Protein kinase phosphorylation site sequences and consensus specificity motifs: Tabulations. in *Methods in Enzymology* 200, 62-81.

Pons, M., and Quintanilla, M. (**2006**). Molecular biology of malignant melanoma and other cutaneous tumors. *Clinical and Translational Oncology*, *8*(7), 466-474.

Rad, E., Murray, J. T., and Tee, A. R. (**2018**). Oncogenic signalling through mechanistic target of rapamycin (mTOR): A driver of metabolic transformation and cancer progression. *Cancers*, *10*(1), 5.

Richards, J. M., Bedikian, A., Gonzalez, R., Atkins, M. B., Whitman, E., Lutzky, J., and Thompson, J. (**2005**). High-dose allovectin-7 in patients with advanced metastatic melanoma: Final phase 2 data and design of phase 3 registration trial. *Journal of Clinical Oncology*, *23*(16), 7543-7543.

Risso, G., Blaustein, M., Pozzi, B., Mammi, P., and Srebrow, A. (**2015**). AKT/PKB:One kinase, many modifications. *Biochemical Journal*, *468*(2), 203-214.

Sacco, F., Perfetto, L., and Cesareni, G. (**2017**). Combining phosphoproteomics datasets and literature information to reveal the functional connections in a cell phosphorylation network. *Proteomics* 18(5-6),1700311.

Sanches, M. M., de Almeida, L. S., and Freitas, J. P. (**2018**). Genes e Melanoma. *Revista da Sociedade Portuguesa de Dermatologia e Venereologia*,*75*(3), 231-238.

Satyamoorthy, K., and Herlyn, M. (**2002**). Cellular and molecular biology of human melanoma. *Cancer Biology and Therapy*, *1*(1), 14-17.

Sen, B., and Johnson, F. M**.** (**2011**). Regulation of SRC family kinases in human cancers. *Journal of Signal Transduction*, 2011,1-14. Siroy, A. E., Davies, M. A., and Lazar, A. J. (**2016**). The PI3K-AKT pathway in melanoma. *Genetics of Melanoma*, 165-180.

Smeenk, L., Van Heeringen, S. J., Koeppel, M., Gilbert, B., Janssen-Megens, E., Stunnenberg, H. G., and Lohrum, M. (**2011**). Role of p53 serine 46 in p53 target gene regulation. *PloS One*, *6*(3), 17574.

Son, Y., Cheong, Y. K., Kim, N. H., Chung, H. T., Kang, D. G., and Pae, H. O. (**2011**). Mitogen-activated protein kinases and reactive oxygen species: how can ROS activate MAPK pathways? *Journal of signal transduction*, *2011*,1-6.

Siroy, A. E., Davies, M. A., and Lazar, A. J. (**2016**). The PI3K-AKT pathway in melanoma. *Genetics of Melanoma*, 165-180.

Steeg, P. S. (**2016**). Targeting metastasis. *Nature Reviews Cancer*, *16*(4), 2016.

Sun, Y., Liu, W. Z., Liu, T., Feng, X., Yang, N., and Zhou, H. F. (**2015**). Signaling pathway of MAPK/ERK in cell proliferation, differentiation, migration, senescence and apoptosis. *Journal of Receptors and Signal Transduction*, *35*(6), 600-604.

Sur, S., and Agrawal, D. K. (**2016**). Phosphatases and kinases regulating CDC25 activity in the cell cycle: Clinical implications of CDC25 over expression and potential treatment strategies. *Molecular and Cellular Biochemistry*, *416*(1-2), 33-46.

Szöőr, Á., Ujlaky-Nagy, L., Tóth, G., Szöllősi, J., and Vereb, G. (**2016**). Cell confluence induces switching from proliferation to migratory signaling by site-selective phosphorylation of PDGF receptors on lipid raft platforms. *Cellular Signalling*, *28*(2), 81-93.

Tai, Y. L., Chen, L. C., and Shen, T. L. (**2015**). Emerging roles of focal adhesion kinase in cancer. *Bio Med Research International*, 2015, 1-13.

Tkáčová, J., and Angelovičová, M. (**2012**). Heat shock proteins (HSPs): A review. *Scientific Papers Animal Science and Biotechnologies*, *45*(1), 349-353.

Vlaeminck-Guillem, V., Gillet, G., and Rimokh, R. (**2014**). SRC: Marker or actor in prostate cancer aggressiveness. *Frontiers in Oncology*, 4, 222.

Völkers, M., and Sussman, M. (**2013**). mTOR/PRAS40 interaction: Hypertrophy or proliferation. *Cell Cycle*,12:23, 3579–3580.

Wee, P., and Wang, Z. (**2017**). Epidermal growth factor receptor cell proliferation signaling pathways. *Cancers*, *9*(5), 52.

Willert, K., and Nusse, R. (**2012**). WNT proteins. *Cold Spring Harbor Perspectives in Biology*, *4*(9), a007864.

Wu, X., and Chen, J. (**2003**). Autophosphorylation of checkpoint kinase 2 at serine 516 is required for radiation-induced apoptosis. *Journal of Biological Chemistry*, *278*(38), 36163-36168.

Xu, W., and McArthur, G. (**2016**). Cell cycle regulation and melanoma. *Current Oncology Reports*, *18*(6), 34.

Xu, Xu, W., Tsvetkov, L. M., and Stern, D. F. (**2002**). Chk2 activation and phosphorylation-dependent oligomerization. *Molecular and Cellular Biology*, *22*(12), 4419-4432.

Zannini, L., Delia, D., and Buscemi, G. (**2014**). CHK2 kinase in the DNA damage response and beyond. *Journal of Molecular Cell Biology*, *6*(6), 442-457.

Zhao, H. F., Wang, J., and Tony To, S. S. (**2015**). The phosphatidylinositol 3-kinase/AKT and c-Jun N-terminal kinase signaling in cancer: Alliance or contradiction*? International Journal of Oncology*, *47*(2), 429-436.

Zhao, X., and Guan, J. L. (**2011**). Focal adhesion kinase and its signaling pathways in cell migration and angiogenesis. *Advanced Drug Delivery Reviews*, *63*(8), 610-615.